\documentclass[12pt]{article}

\usepackage{latexsym,amsmath,amscd,amssymb,amsthm,graphics}
\usepackage{enumerate,color}
\usepackage[margin=2cm]{geometry}
\usepackage{graphicx}
\usepackage[colorlinks]{hyperref}
\usepackage{url}
\usepackage{framed}
\usepackage[all]{xy}

\DeclareMathOperator{\curl}{curl}

\DeclareMathOperator{\diff}{d\!}
\DeclareMathOperator{\dx}{\diff{x}}%{\diff\!\mathbf{x}}

\DeclareMathOperator{\SSA}{SSA}

\DeclareMathOperator{\SP}{SP}

\def\MM#1{\boldsymbol{#1}}
\newcommand{\pp}[2]{\frac{\partial #1}{\partial #2}} 
\newcommand{\dede}[2]{\frac{\delta #1}{\delta #2}}
\newcommand{\dd}[2]{\frac{\diff #1}{\diff #2}}

\newcommand{\bfi}{\bfseries\itshape}
\newcommand{\rem}[1]{}

\numberwithin{equation}{section}
\numberwithin{figure}{section}

\makeatother

\begin{document}

\title{Variational formulations of sound-proof models} 
\author{C. J. Cotter$^{1}$ and D. D. Holm$^{2}$}
\addtocounter{footnote}{1}
\footnotetext{Department of Aeronautics, Imperial College London. London SW7 2AZ, UK. 
\texttt{colin.cotter@imperial.ac.uk}
\addtocounter{footnote}{1} }
\footnotetext{Department of Mathematics, Imperial College London. London SW7 2AZ, UK. Partially supported by European Research Council Advanced Grant 267382.
\texttt{d.holm@imperial.ac.uk}
\addtocounter{footnote}{1} }

\date{}
\maketitle

\makeatother

\maketitle

\noindent \textbf{AMS Classification:} 


\begin{abstract}
  We derive a family of ideal (nondissipative) 3D sound-proof fluid models that includes both the Lipps-Hemler anelastic approximation (AA) and the Durran pseudo-incompressible approximation (PIA). This family of models arises in the Euler-Poincar\'{e} framework involving a constrained Hamilton's principle expressed in the Eulerian fluid description. The derivation in this framework establishes the following properties of each member of the entire family: the Kelvin-Noether circulation theorem, conservation of potential vorticity on fluid parcels, a Lie-Poisson Hamiltonian formulation possessing conserved Casimirs, a conserved domain integrated energy and an associated variational principle satisfied by the equilibrium solutions. \smallskip
  
Having set the stage with the derivations of 3D models using the constrained Hamilton's principle, we then derive the corresponding 2D vertical slice models for these sound-proof theories. \\

\end{abstract}

\maketitle

%%%%%%%%%%%%%%%%%%%%%%%%%%%%%%%%%%%%%%%%%%%%%%%%%%%%%%%%%%%%%%%%
\section{Introduction} \label{sec-intro}

The fastest-moving atmospheric and oceanic waves are the sound waves, whose presence can adversely affect numerical simulations of
atmospheric and ocean circulations, by poisoning the desired low
frequency circulations with high frequency oscillations. These high
frequency oscillations must somehow be filtered or removed in order to facilitate numerical simulations of the slower motions that are needed in predictions of weather and climate. An effective way of
\emph{sound-proofing} the numerical simulations is to replace the
exact governing equations with an approximate system that does not
possess sound waves. Leading candidates for the sound-proofed
approximations include the anelastic approximation (AA) of
\cite{OgPh1962}, \cite{LH1982} and \cite{Ba1996}, and the
pseudo-incompressible approximation (PIA) of 
\cite{Du1989,Du2008}. Recently,  \cite{Kl2009} re-examined the
range of scales for which the AA and PIA sound-proof fluid
approximations are valid by using the method of asymptotic expansions.

Here we derive a family of ideal (nondissipative) 3D sound-proof fluid models 
from a constrained Hamilton's principle in the Euler-Poincar\'e framework. 
The  AA and PIA arise in this family from applying different constraints. 
We then pursue the implications of the framework, particularly for
conservation of energy and potential vorticity. After analysing the 
basic properties of this 3D family of sound-proof models, we derive their
corresponding 2D slice models and analyse their properties once again
using the Euler-Poincar\'e approach recently developed in
\cite{CoHo2012}. These slice models assume a $y$-independent solution
structure but include the $y$-component of velocity and the Coriolis
force. 
Because they assume a constant $y$-gradient of potential temperature, 
they possess solutions that are not
solutions of the full three
dimensional equations. Slice models have been used to study the
formation and subsequent evolution of fronts, and are very useful for
benchmarking numerical schemes since they can be run quickly on a
single workstation \cite{Cu2007}. 

%%%%%%%%%%%%%%%%%%%%%%%%%%%%%%%
\subsection{Anelastic approximation (AA)}
 \cite{Ba1996} examined the anelastic approximation for deep
fluid convection and proposed an alternative form of the anelastic
equations. This alternative model combines the results of  \cite{DF1969} and \cite{LH1982} to produce a hybrid theory that 
(1) conserves the domain integrated energy; 
(2) preserves potential vorticity on fluid parcels; and 
(3) accurately represents the acoustic adjustment process in Lamb's problem, cf.  \cite{Ba1995}. These equations for a dry anelastic compressible
fluid (atmosphere) rotating at angular frequency $\boldsymbol{\Omega}$
under constant vertical gravitational acceleration $g\hat{\bf z}$
comprise the following constrained dynamical system,
\begin{eqnarray}
\frac{d\mathbf{u}}{dt} + 2 \boldsymbol{\Omega} \times \mathbf{u}
&=&
-\ \nabla\bigg(\frac{p'}{\rho_0}\bigg)
+ \frac{g\theta'}{\theta_0}\hat{\bf z}
\,,
\label{AA-mot-eqn}\\
\frac{d(\theta_0+\theta')}{dt}&=&0
\,,\label{entropy-advect-eqn}
\end{eqnarray}
{with} the density-weighted incompressibility constraint 
\begin{align}
\nabla\cdot(\rho_0(z)\mathbf{u})=0
\,,
\label{AA-constraint}
\end{align}
and two diagnostic relations for the thermodynamic variables $\rho'$ and $T'$,
\begin{align}
\frac{\theta'}{\theta_0}
=
\frac{p'}{\rho_0 g H_{\rho}}
- \frac{\rho'}{\rho_0}
\,,\label{diagnostic-rel}
\quad\hbox{and}\quad
\frac{p'}{p_0}
=
\frac{\rho'}{\rho_0} + \frac{T'}{T_0}
\,.
\end{align}
In these equations, the fluid velocity is denoted $\mathbf{u}$, the
advective time derivative is written as
$d/dt=\partial/\partial{t}+\mathbf{u}\cdot\nabla$, and the state
variables for this anelastic motion are: pressure $p$, density $\rho$,
potential temperature $\theta$, and temperature $T$. Each of these
state variables consists of the sum of the base state (with subscript
$0$) and a dynamic contribution denoted with a prime, as in
\begin{eqnarray*}\label{theta-decomp}
\theta(x,y,z,t) &=& \theta_0(z) + \theta'(x,y,z,t)\,,
\\
D(x,y,z,t) &=& \rho_0(z) + \rho'(x,y,z,t)\,,
\quad\hbox{etc.},
\end{eqnarray*}
where $\theta'$ (resp. $\rho'$) is the dynamic contribution to the potential temperature (resp. mass density) field. The base state satisfies
\begin{equation}\label{base-rel}
\frac{dp_0}{dz}=-g\rho_0
\,,\quad
p_0=\rho_0 R T_0
\,,\quad
C_v\,\theta_0\frac{d\pi_0}{dz}=-\,g
\,,
\end{equation}
where $\pi_0=T_0/\theta_0$. The constants $R$ and $C_v$ are the
ideal gas constant $R$ and the specific heat at constant pressure
$C_v$ for dry air. The scale height of the base density state is 
\begin{equation}\label{dens-scale-ht}
1/H_{\rho}=-\rho_0(z)^{-1}d\rho_0/dz
\,.
\end{equation}
Given the base state
functions satisfying relations (\ref{base-rel}), as well as the
velocity $\mathbf{u}$ and the dynamic contributions
$p'$ and $\theta'$ at any time, the system is completed by the constraint in (\ref{AA-constraint}),  whose preservation determines $p'$, and the thermodynamic diagnostic relations in (\ref{diagnostic-rel}). The distinctions between the anelastic approximation (AA) and the traditional models {}\cite {DF1969}, {}\cite {LH1982} are discussed in detail by \cite{Ba1996} and \cite{Kl2009}. The important point for us here is that the model equations
(\ref{AA-mot-eqn}) - (\ref{dens-scale-ht}) agree in both formulations, that of \cite{LH1982} and that of \cite{Ba1996}.

%%%%%%%%%%%%%%%%%%%%%%%%%%%%%%%
\subsection{Pseudo-incompressible approximation (PIA)}
The dynamical equations of the pseudo-incompressible approximation (PIA) are given by \cite{Du1989} as a motion equation and an advection law. These are, respectively, 
\begin{eqnarray}
&&\frac{d\mathbf{u}}{dt} + 2 \boldsymbol{\Omega} \times \mathbf{u}
=
-\ (\theta_0+\theta')\nabla{p'}
+ \frac{g\theta'}{\theta_0}\hat{\bf z}
\,,
\label{PIA-mot-eqn}\\
&&\hbox{and}\quad
\frac{d(\theta_0+\theta')}{dt}=0
\,.
\label{PIA-advect-eqn}
\end{eqnarray}
Note that the pressure force in the PIA motion equation \eqref{PIA-mot-eqn} differs from that in the AA motion equation \eqref{AA-mot-eqn}.
The weighted incompressibility constraint for PIA is
\begin{equation}
\nabla\cdot\big(\rho_0(z)\,\theta_0(z)\mathbf{u}\big) 
=0
\,,
\label{PIA-constraint}
\end{equation}
which also differs from the corresponding density-weighted incompressibility constraint \eqref{AA-constraint} for AA.
However, both the entropy (i.e., heat and buoyancy) advection law \eqref{PIA-advect-eqn} and the two diagnostic relations (\ref{diagnostic-rel}) for the thermodynamic variables are the \emph{same} for the two models AA and PIA.

The AA and PIA equations differ in the pressure forces in their motion
equations, \eqref{AA-mot-eqn} versus \eqref{PIA-mot-eqn}, and in their
weighted incompressibility constraints, \eqref{AA-constraint} versus
\eqref{PIA-constraint}. As we shall see, these differences are
related. Namely, the difference in pressure forces arises from a
difference in the constraints enforced by the pressure as a
Lagrange multiplier, in the Hamilton's principles for the two
theories.

\subsection{Outlook} 

In this paper, we show that both sets of motion equations
(\ref{AA-mot-eqn}) for AA and (\ref{PIA-mot-eqn}) for PIA are members 
of a single family of Euler-Poincar\'{e} equations obtained from a
constrained Hamilton's principle expressed in the Eulerian fluid
description. Two sets of constraints arise from the underlying Lagrangian fluid flow.
One set of constraints affects the variations of velocity, density and potential temperature. 
The other set of constraints imposes sound-proofing by using the pressure as 
a Lagrange multiplier to enforce conditions on the density and potential temperature. The 
Euler-Poincar\'{e} formulation of this family of sound-proofed fluid equations
provides their Kelvin-Noether circulation theorems which lead to the
conservation of potential vorticity on fluid parcels.
Domain-integrated energy is also conserved for the entire sound-proofed family of equations.
This result can be demonstrated by applying a Legendre transformation to obtain a
Lie-Poisson bracket formulation of the sound-proofed family of equations. 
In turn, the Lie-Poisson Hamiltonian formulation reveals a set of
conserved integral quantities known as Casimirs. 
In the underlying Lagrangian fluid flow, the Casimirs generate motion of the fluid
along streamlines of the steady solutions of the sound-proofed family of equations.
Thus, the Casimirs provide a constrained-energy variational principle for 
obtaining equilibrium solutions of sound-proofed fluid dynamics. They also form the basis for
determining the Lyapunov stability conditions for these equilibria, following the
approach applied to the Euler-Boussinesq equations in \cite{AHMR1986}. 
(The Euler-Boussinesq equations form a special
case of the sound-proofed models in which the base state is constant.)
In general, these conservation laws are very important for the
analysis of geophysical flows and the design of numerical schemes for
weather prediction and climate modelling. For example, in
\cite{Cu2007}, conservation laws were emphasised in analysing the
ability of numerical schemes to reproduce balanced model solutions in
asymptotic limits. See \cite{Th2008} for a review of the relevance of
conservation laws to the design of dynamical cores of weather and
climate models. The Euler-Poincar\'{e} formulation allows us to
produce hierarchies of models that all share the same conservation
properties, from fully compressible deep atmosphere models to
semi-geostrophic balanced models, for example. Here we show that the 
Euler-Poincar\'{e} formulation also unifies the family of sound-proofed equations. 

\paragraph{Previous related work.}
In previous related work on Hamiltonian structures of soundproof
models, a two-dimensional study of the Hamiltonian structure of the
Lipps-Hemler AA model was presented by \cite{SS1992}, who also studied wave-activity conservation laws in the two-dimensional case. 
A canonical Hamiltonian formulation of the 3D Lipps-Hemler AA
model in the Lagrangian fluid description was given in Appendix A of 
\cite{Be1995}.  As one would
expect, the Lie-Poisson Hamiltonian formulation for the Eulerian
fluid description of these equations is equivalent to the canonical
formulation of \cite{Be1995} in the Lagrangian fluid description.
This equivalence is guaranteed by the Euler-Poincar\'e theorem {}\cite {HMR1998a}.
This theorem was developed for applications
in geophysical fluid dynamics (GFD) in \cite{HMR1998b} and specialised to
slice geometries in \cite{CoHo2012}. It states that the following four
dynamical perspectives of fluid mechanics are equivalent: (1)
Hamilton's principle for the Lagrangian fluid description; (2) the
Euler-Lagrange equations in the Lagrangian fluid description; (3)
Hamilton's principle for the Eulerian fluid description with certain
constrained variations; and (4) the Euler--Poincar\'e equations in the
Eulerian fluid description.

The methods of this paper are based on the Euler--Poincar\'e theory of
reduction by symmetry of variational principles introduced in \cite{HMR1998a}. However, in this paper we shall just quote the main results of this theory and apply them to derive a family of
sound-proofed models.

\paragraph{Organization of the Paper.} In \S\ref{sec-EP} we recall
from \cite{HMR1998a,HMR1998b} the results of
the Euler-Poincar\'{e} theorem for Lagrangians in continuum mechanics
depending on advected quantities (such as density and potential
temperature), along with their associated Kelvin--Noether theorem and
Lie-Poisson Hamiltonian formulations. These results establish the
mathematical framework into which we place the dynamical equations for
the AA and PIA models in \S\ref{sec-anelastic-eqns}.  In
\S\ref{vertsliceAA+PIA-sec} we introduce soundproof slice models that
preserve the fluid properties guaranteed by the Euler-Poincar\'{e}
framework. Summary conclusions of the main results of the paper and
some of the outstanding problems it raises are discussed in
\S\ref{conc-sec}.

%%%%%%%%%%%%%%%%%%%%%%%%%%%%%%%%%%%%%%%%%%%%%%%%%%%%%%%%%%%%%%%%%
\section{The Euler--Poincar\'{e} Theorem in GFD}
\label{sec-EP}

We begin by recalling from \cite{HMR1998a,HMR1998b} the statements of the Euler--Poincar\'e
equations and their associated Kelvin--Noether theorem in the context
of continuum mechanics and approximate models in GFD. 

The Euler-Poincar\'{e} equations for a GFD reduced Lagrangian
$\ell\,[\mathbf{u},D,\theta\,]$ (usually the domain integrated kinetic
energy minus the domain integrated potential energy) involve the Eulerian  fluid
velocity vector $\mathbf{u}$, the scalar potential temperature
$\theta$ and the mass density $D$ as functions of three dimensional space
with coordinates $\mathbf{x}$ and time $t$. In vector notation, these
equations are expressed as \cite{HMR1998a,HMR1998b},
\begin{equation}
\frac{d}{dt} \frac{1}{D} \frac{{\delta} \ell}{{\delta} \mathbf{u}}
\,+\, \frac{1}{D} \frac{{\delta} \ell}{{\delta} u^j} \nabla u^j
\,+\, \frac{1}{D} \frac{{\delta} \ell}{{\delta} \theta} \nabla \theta
-  \nabla\frac{{\delta} \ell}{{\delta} D}=0,
\label{EP-comp1}
\end{equation}
or, equivalently, in \emph{curl form} as,
\begin{align}
\nonumber
0 & =  \frac{\partial}{\partial t}
\Big(\frac{1}{D} \frac{{\delta} \ell}{{\delta} \mathbf{u}}\Big)
\,-\, \mathbf{u}\times {\rm curl}
\Big(\frac{1}{D} \frac{{\delta} \ell}{{\delta} \mathbf{u}}\Big)  \\
&  \quad \,+\, \nabla\Big(\mathbf{u}\cdot\frac{1}{D}
\frac{{\delta} \ell}{{\delta} \mathbf{u}}
\,-\, \frac{{\delta} \ell}{{\delta} D}\Big)
\,+\, \frac{1}{D} \frac{{\delta} \ell}{{\delta} \theta} \nabla \theta,
\label{EP-comp2}
\end{align}
where $\delta l/\delta D$ is the \emph{variational derivative} of $l$
with respect to $D$, defined by
\[
\int\frac{{\delta} \ell}{{\delta} D}\delta D\diff x
= \lim_{\epsilon\to 0} \frac{\ell[\mathbf{u},D+\epsilon \delta D,\theta]-
\ell[\mathbf{u},D,\theta]}{\epsilon},
\]
for all density perturbations $\delta D$. Similar definitions hold for
$\delta l/\delta\theta$ and $\delta l/\delta\mathbf{u}$ (the latter of
which is the linear momentum, a vector quantity).
The Euler--Poincar\'e system is completed by including the
\emph{auxiliary equations} for advection of the total potential
temperature $\theta$,
\begin{equation}
\frac{\partial{\theta}}{\partial{t}}+\mathbf{u}\cdot\nabla{\theta} 
=
0\,,
\label{theta-advect}
\end{equation}
and the continuity equation for the mass density $D$,
\begin{equation}
\frac{\partial{D}}{\partial{t}}
+\nabla\cdot({D}\mathbf{u}) 
=
0\,.
\label{D-cont}
\end{equation}
Specific models are then obtained (for example, the sound-proof models
discussed in this paper) by choosing a particular form of the
Lagrangian $\ell$.

%%%%%%%%%%%%%%%%%%%%%%%%%%%%%%%%%%
%\vspace{1in}
%For incompressible, anelastic and pseudoincompressible flows, the
%pressure plays the role of a Lagrange multiplier in Hamilton's
%principle that enforces a constraint on the mass density $D$. The 
%constraints for each model are all members of the same family:
%\begin{itemize}
%\item
%For incompressible flows, one sets $D=1$ in the continuity equation  \eqref{D-cont}, so that 
%\[
%\nabla \cdot \mathbf{u} = 0
%\,.
%\]
%\item
%For AA flows, one sets $D=\rho_0(z)$ in the continuity equation \eqref{D-cont} with a prescribed stably stratified reference density profile $\rho_0(z)$, so that
%\[
%\nabla\cdot(\rho_0(z)\mathbf{u})=0
%\,.
%\]
%\item
%Finally, for PIA flows, one combines \eqref{theta-advect} and \eqref{D-cont} into
%\begin{equation}
%\frac{\partial{(D\theta)}}{\partial{t}}
%+\nabla\cdot({D\theta}\mathbf{u}) 
%=
%0\,,
%\label{thetaD-cont}
%\end{equation}
%for which the constraint $D\,\theta = \rho_0(z)\,\theta_0(z)$ imposed in the Hamilton's principle for PIA implies 
%\[
%\nabla\cdot(\rho_0(z)\,\theta_0(z)\mathbf{u})=0
%\,.
%\]
%\end{itemize}
%%
%%%%%%%%%%%%%%%%%%%%%%%%%%%%%%%%%%

\paragraph{Kelvin-Noether circulation theorem.}
The Euler--Poincar\'e motion equation in either form
(\ref{EP-comp1}) or (\ref{EP-comp2}) results in the {\bfi
Kelvin-Noether circulation theorem},
\begin{equation} \label{KN-theorem-bD}
\frac{d}{dt}\oint_{\gamma_t(\mathbf{u})} \frac{1}{D}\frac{\delta
\ell}{\delta \mathbf{u}}\cdot d\mathbf{x}
= -\oint_{\gamma_t(\mathbf{u})}
\frac{1}{D}\frac{\delta \ell}{\delta \theta}\nabla \theta \cdot d\mathbf{x}\;,
\end{equation}
where the curve $\gamma_t(\mathbf{u})$ moves with the fluid velocity
$\mathbf{u}$. Then, by Stokes' theorem, the Euler--Poincar\'e
equations generate circulation of the quantity
$D^{-1}{\delta{\ell}/\delta\mathbf{u}}$ whenever the
gradients $\nabla \theta$ 
and $\nabla(D^{-1}\delta \ell/\delta{\theta})$ are not collinear.

\paragraph{Potential vorticity conservation laws.}
Taking the curl of equation (\ref{EP-comp2}) and using advection of the potential temperature $\theta$ and the continuity equation for the density $D$ yields {\bfi conservation of potential vorticity PV on fluid
parcels}, as expressed by
\begin{align} \label{pv-cons-EP}
\frac{\partial{q}}{\partial{t}}+\mathbf{u}\cdot\nabla{q} &=
0\,, \\
\hbox{where}\quad
{q} &= \frac{1}{D}\nabla{\theta}\cdot{\rm curl}
\left(\frac{1}{D}\frac{\delta \ell}{\delta \mathbf{u}}\right).
\end{align}
Consequently, the following domain integrated quantities are
conserved, for any smooth function $\Phi$,
\begin{equation} \label{EP-Casimirs}
C_{\Phi} = \int d^{\,3}x\ 
D\,\Phi(\theta,q)\,,
\quad\forall\,\Phi\,.
\end{equation}
\paragraph{Legendre transform, energy conservation and Hamiltonian formulation.}

The absence of explicit time dependence in the Lagrangian
$\ell\,[\mathbf{u},D,\theta\,]$ gives the {\bfi conserved domain integrated energy}, via Noether's theorem for time translation invariance. This energy is easily calculated using the {\bfi Legendre transform} to be
\begin{equation} \label{EP-erg}
E\,[\mathbf{u},D,\theta\,] = \int d^{\,3}x\ 
\Big(\mathbf{u}\cdot
\frac{\delta \ell}{\delta \mathbf{u}}\Big)
- \ell\,[\mathbf{u},D,\theta\,]
\,.
\end{equation}
When the Legendre transform is completed to express
$E\,[\mathbf{u},D,\theta\,]$ as $H\,[\mathbf{m},D,\theta\,]$ with
\begin{equation}\label{m-def}
\mathbf{m}\equiv\delta \ell/\delta \mathbf{u}
\quad\hbox{and}\quad
\delta{H}/\delta\mathbf{m}=\mathbf{u}
\,,
\end{equation}
the Euler--Poincar\'e system (\ref{EP-comp1})--(\ref{D-cont}) may be expressed in Hamiltonian form as
\begin{equation}\label{mu-dot-syst}
\frac{\partial \mu}{\partial t}=\{\mu,H\}\,,
\quad \hbox{with}\quad
\mu\in[\mathbf{m},D,\theta\,]\,,
\end{equation}
and the {\bfi Lie-Poisson bracket} is given in Euclidean component form by
\begin{align} 
\begin{split}
& \{F,H\}[\mathbf{m},D,\theta\,] =
\\ 
& \ - \, \int d^{\,3}x\ \bigg\{
\sum_i\frac{\delta F}{\delta m_i}
\bigg[\sum_j\big(\partial_j m_i+m_j \partial_i\,\big)
\frac{\delta H}{\delta m_j} \\
& \hspace{2cm} 
\ +\ \big(D\partial_i\,\big)\frac{\delta H}{\delta D}
\ -\ \big( \theta_{,\,i}\,\big)\frac{\delta H}{\delta \theta}\bigg]
\\
&\hspace{1cm}
+\ \frac{\delta F}{\delta D}\sum_j\big(\partial_j D\big)
\frac{\delta H}{\delta m_j}
\ +\ \frac{\delta F}{\delta \theta}
\sum_j\big( \theta_{,\,j}\big)
\frac{\delta H}{\delta m_j}
\bigg\}\,.
\end{split}
\label{LPB}
\end{align}
The conserved quantities $C_{\Phi}$ in (\ref{EP-Casimirs}) are then
understood in the {\bfi Lie-Poisson Hamiltonian formulation}
(\ref{mu-dot-syst}) -- (\ref{LPB}) of the Euler--Poincar\'e system
(\ref{EP-comp1}) -- (\ref{D-cont}) as {\bfi Casimirs} that commute
under the Lie-Poisson bracket (\ref{LPB}) with any functional of
$(\mathbf{m},D,\theta\,)$. The Casimirs also result via Noether's
theorem from symmetry of the Hamilton's principle for the
Euler--Poincar\'e system under the \emph{particle relabeling
  transformations} that leave invariant the Lagrangian
$\ell[\mathbf{u},D,\theta\,]$.  For full mathematical details,
consult \cite{MR1994,HMR1998a,HMR1998b,CoHo2012b}.
\paragraph{The Euler--Poincar\'e framework.}
The four properties (\ref{KN-theorem-bD})--(\ref{EP-erg}) and the
Lie-Poisson Hamiltonian formulation (\ref{mu-dot-syst}) -- (\ref{LPB})
of the Euler--Poincar\'e equation (\ref{EP-comp1}) and its auxiliary
equations (\ref{theta-advect}) and (\ref{D-cont}) are all desirable
elements of approximate models for applications in geophysical fluid
dynamics expressed in the variables $(\mathbf{u},D,\theta\,)$. Thus,
the Euler--Poincar\'e theory offers a unified framework in which to
derive approximate GFD models that possess these properties: the
Kelvin-Noether circulation theorem, conservation of potential
vorticity on fluid parcels, and the Lie-Poisson Hamiltonian
formulation with its associated conserved Casimirs and conserved
domain integrated energy. Previous work {}\cite {HMR1998a,HMR1998b} has
shown that many useful GFD approximations may be formulated as
Euler--Poincar\'e equations, whose shared properties thus follow from
this underlying common framework. The aim of the next section of this
paper is to cast the sound-proof motion equations, including those for 
AA (\ref{AA-mot-eqn}) and for PIA (\ref{PIA-mot-eqn}), into the 
Euler--Poincar\'e framework.

%%%%%%%%%%%%%%%%%%%%%%%%%%%%%%%%%%%%%%%%%%%%%%%
\section{Hamilton's principle for sound-proof motion equations}
\label{sec-anelastic-eqns}

\subsection{The sound-proof Lagrangians} 
In the Eulerian fluid representation, we consider Hamilton's principle
$\delta S = 0$ for fluid motion in a three dimensional domain with action functional
${\cal S}=\int\,dt\, \ell$ and Lagrangian
$\ell[\mathbf{u},D,\theta\,]$. 
For the sound-proof models, we propose a Lagrangian 
given by the total kinetic energy in the rotating frame, minus the total potential energy
including the thermodynamic energy, 
plus the sound-proof constraint imposed by the pressure $p'$ as a Lagrange multiplier,
\begin{align} 
\nonumber
\ell_{\SP} &= \int \bigg[ D
\bigg(\frac{1}{2} |\mathbf{u}|^2
+ \mathbf{u}\cdot\mathbf{R}(\mathbf{x}) - gz
- C_v\pi_0(z)\,\theta\bigg) \\
&  \qquad +\
{p'}
\underbrace{
\Big(
 \rho_0(z)\varTheta(\theta_0(z)) - D\,\varTheta(\theta)\Big)
}_{\hbox{SP constraint}}\
\bigg]d^{\,3}x
\,.
\label{SP-lag}
\end{align}
Here $D$ denotes the mass density, $\mathbf{u}$ is the Eulerian fluid velocity,
$\mathbf{R}$ is a vector field whose curl is $2\boldsymbol{\Omega}$ 
(twice the local angular rotation vector), 
$\theta$ is the total potential temperature, 
$\theta_0(z)$, $\pi_0(z)$ and $\rho_0(z)$ are reference profiles and
$\varTheta$ is an arbitrary smooth function still to be chosen. 
\paragraph{The constraint imposed by the pressure.}
In the Lagrangian \eqref{SP-lag} the pressure $p'$ plays the role of a Lagrange multiplier in Hamilton's
principle that enforces a constraint that relates the mass density $D$ and potential temperature $\theta$. To understand the meaning of this constraint, one may combine \eqref{theta-advect} and \eqref{D-cont} into
\begin{equation}
\frac{\partial (D\varTheta(\theta))}{\partial{t}}
+\nabla\cdot({D\varTheta(\theta)}\mathbf{u}) 
=
0\,,
\label{thetaD-cont}
\end{equation}
where $\varTheta$ is a smooth function. Imposing the sound-proofing constraint $D\,\varTheta(\theta) = \rho_0(z)\,\varTheta(\theta_0(z))$ then yields
\begin{equation}
\nabla\cdot\big(\rho_0(z)\,\varTheta(\theta_0(z))\mathbf{u}\big) 
=
0\,.
\label{SP-constraint1}
\end{equation}
For $\varTheta(\theta)=1-\alpha + \alpha \theta$, one recovers PIA flows for $\alpha=1$ and AA flows for $\alpha=0$. For $\alpha=0$ and $\rho_0(z)=const$, one recovers divergence-free flows. 

\paragraph{Time-dependent reference states.}
Following \cite{Du2008} one may allow reference states with \emph{prescribed} time-dependence by replacing $\rho_0(z)$ by $\widetilde{\rho}(x,y,z,t)$, $\theta_0(z)$ by $\widetilde{\theta}(x,y,z,t)$ and $\pi_0(z)$ by $\widetilde{\pi}(x,y,z,t)$. Imposing the sound-proofing constraint $D\,\varTheta(\theta) = \widetilde{\rho}\,\varTheta(\widetilde{\theta})$ then yields
\begin{equation}
\nabla\cdot\big(\widetilde{\rho}\,\varTheta(\widetilde{\theta})\mathbf{u}\big) 
=
-\,\partial_t (\widetilde{\rho}\,\varTheta(\widetilde{\theta}))
\,,
\label{tilde-cont}
\end{equation}
in which the right-hand side is a prescribed function of $(x,y,z,t)$. 
The pressure is then determined from the sound-proof motion equation via an elliptic equation involving
\[
\nabla\cdot \left(\widetilde{\rho}\,\varTheta(\widetilde{\theta})\partial_t \mathbf{u}\right)
=
- \,\partial_t^2 \left(\widetilde{\rho}\,\varTheta(\widetilde{\theta})\right)
- \nabla\cdot \left(\mathbf{u}\partial_t \big(\widetilde{\rho}\,\varTheta(\widetilde{\theta})\big)\right)
.\]
The resulting sound-proof fluid equations keep their form, as long as the time-dependent background state remains in hydrostatic balance. However, the time-dependence in $\widetilde{\rho}$, $\widetilde{\theta}$ and $\widetilde{\pi}$ prevents conservation of energy, although potential vorticity is still conserved on fluid parcels. 
%In particular, we find the energy relation
%\begin{equation} 
%\frac{d E_{sp}}{dt} = \int 
%DC_v\theta\, \partial_t \tilde{\pi} - p' \partial_t  \left(\widetilde{\rho}\,\varTheta(\widetilde{\theta})\right)
%d^3x
%\,.
%\label{sp-erg-time}
%\end{equation}for the total sound-proof energy $E_{sp}$ given as
%\begin{equation} \label{sp-erg-tilde}
%E_{\SP} = \int 
%\bigg(\frac{1}{2} |\mathbf{u}|^2
%+ gz
%+ C_v\widetilde{\pi}\theta\bigg)
%D \,d^{\,3}x
%\,.
%\end{equation}

%

%%%%%%%%%%%%%%%%%%%%%%%%%%%%%%%%%%%%%%%%%%%%
%The
%corresponding Lagrangian for PIA is
%
%The only difference between the Hamilton's principles for the AA and
%PIA models is in their constraints, imposed by the corresponding
%pressures $p'$ as Lagrange multipliers.
%
%%
%The two Lagrangians $\ell_{\AAm}$ and $\ell_{\PIA}$ possess the following variational derivatives at fixed $\mathbf{x}$ and $t$. For $\ell_{\AAm}$ we have
%%
%\begin{align}
%\begin{split}
%&\frac{1}{D}\frac{{\delta} \ell_{\AAm}}{{\delta} \mathbf{u}}
%= \mathbf{u}+ \mathbf{R}(\mathbf{x})\,, \\
%&
%\frac{1}{D} \frac{{\delta} \ell_{\AAm}}{{\delta} \theta}
%= - C_v\pi_0(z)\,,
%\quad
%\frac{{\delta} \ell_{\AAm}}{{\delta} p'} = 1-\frac{D}{\rho_0(z)}\,,
%\\
%&\frac{{\delta} \ell_{\AAm}}{{\delta} D}
%= \frac{1}{2} |\mathbf{u}|^2
%+ \mathbf{u}\cdot\mathbf{R}(\mathbf{x}) - gz
%- C_v\pi_0(z)\theta - \frac{p'}{\rho_0(z)}\,.
%\end{split}
%\label{vds-AA}
%\end{align}
%
%%%%%%%%%%%%%%%%%%%%%%%%%%%%%%%%%%%%%%%%%%%%

\paragraph{Variational derivatives of the Lagrangian $\ell_{\SP}$.}
The Lagrangian $\ell_{\SP}$ 
\color{blue}
in \eqref{SP-lag}
\color{black}
possesses the following variational derivatives at fixed $\mathbf{x}$ and $t$.
\begin{align}
\begin{split}
&\frac{1}{D}\frac{{\delta} \ell_{\SP}}{{\delta} \mathbf{u}}
= \mathbf{u}+ \mathbf{R}(\mathbf{x})
\,,\\
&\frac{{\delta} \ell_{\SP}}{{\delta} D}
= \frac{1}{2} |\mathbf{u}|^2
+ \mathbf{u}\cdot\mathbf{R}(\mathbf{x}) - gz
\\&\hspace{15mm}
- C_v\pi_0(z)\theta - {p'} \varTheta(\theta)
\,,\\
&
\frac{1}{D} \frac{{\delta} \ell_{\SP}}{{\delta} \theta}
= -\, C_v\pi_0(z) -p' \varTheta'(\theta)
\,,\\
&
\frac{{\delta} \ell_{\SP}}{{\delta} p'} 
=  \rho_0(z)\varTheta(\theta_0(z)) - D\,\varTheta(\theta)
\,.
\end{split}
\label{vds-PIA}
\end{align}
%%%%%%%%%%%%

One obtains the corresponding motion equations upon substitution of these variational derivatives into the Euler-Poincar\'e formula (\ref{EP-comp1}). 

\subsection{The motion equation for SP} 
From the Euclidean component formula (\ref{EP-comp2}) for
Hamilton's principles of this type, we find the fluid motion equation for the SP model in three dimensions,
\begin{equation}
\frac{d\mathbf{u}}{dt} 
- \mathbf{u} \times {\rm curl} \mathbf{R}
+\varTheta(\theta) \nabla {p'}
+ \bigg(g + C_v \theta \frac{d\pi_0}{dz}\bigg)\hat{\bf z}
=0
\,,
\label{EPSP-mot}
\end{equation}
where ${\rm curl}\,\mathbf{R}=2\boldsymbol{\Omega}(\mathbf{x})$ is the Coriolis parameter (i.e., twice the local angular rotation
frequency). We use equation (\ref{base-rel}) to rewrite the last  
term in parentheses as
\begin{equation}
g + C_v \theta \frac{d\pi_0}{dz} 
= g\bigg(1-\frac{\theta}{\theta_0}\bigg)
= - \,g \frac{\theta'(\mathbf{x},t)}{\theta_0(z)}\,.
\label{reln1}
\end{equation}
Consequently, we obtain the SP motion equation, namely,
\begin{equation}
\frac{d\mathbf{u}}{dt} 
- \mathbf{u} \times 2\boldsymbol{\Omega}(\mathbf{x})
+\varTheta(\theta) \nabla {p'}
- \frac{g\theta'}{\theta_0}\hat{\bf z}
=0
\,,
\label{SP-mot-eqn1}
\end{equation}
as the Euler--Poincar\'e equation for the SP Lagrangian (\ref{SP-lag}).

\paragraph{Solving for the SP pressure.}
The SP contraint with arbitrary smooth function $\varTheta$
\begin{equation}
 \rho_0(z)\varTheta(\theta_0(z)) = D\,\varTheta(\theta)
\label{SP-constraint2}
\end{equation}
is obtained from stationarity of the Lagrangian (\ref{SP-lag}) with respect to variations in $p'$. 
Upon substituting this constraint into the continuity equation (\ref{thetaD-cont}), we find the SP divergence condition in equation (\ref{SP-constraint1}). 
Preservation of this condition
determines the pressure $p'$ analytically, by solving the elliptic equation obtained by taking the divergence of the SP motion equation \eqref{SP-mot-eqn1} after first multiplying it by the product $\rho_0(z)\varTheta(\theta_0(z))$. 

The boundary condition for the resulting elliptic equation is obtained from the normal component of the SP motion equation \eqref{SP-mot-eqn1} evaluated on the boundary and using the boundary condition for the velocity, e.g., that it has no normal component at the boundary, which yields a Neumann boundary condition for obtaining the pressure.

\subsection{The Kelvin--Noether theorem for SP equations} 
From equation (\ref{KN-theorem-bD}), the Kelvin--Noether circulation theorem corresponding to the SP fluid motion equation (\ref{SP-mot-eqn1}) in three dimensions is,
\begin{align}
\begin{split}
&\frac{d}{dt}\oint_{\gamma_t(\mathbf{u})}(\mathbf{u}
+\mathbf{R})\cdot d\mathbf{x}
\\&\hspace{5mm}
= -\oint_{\gamma_t(\mathbf{u})}\!\!
\big(\varTheta(\theta)\, \nabla p' - \theta\frac{g}{\theta_0} \nabla z\big)\cdot d\mathbf{x}\;,
\\&\hspace{5mm}
= -\int\!\!\!\!\int_{S}\!\!
\nabla \theta\times (\varTheta'(\theta)\nabla p' - \frac{g}{\theta_0} \nabla z)\cdot d\mathbf{x}\;,
\end{split}
\label{KN-theorem-SP}
\end{align}
where the curve $\gamma_t(\mathbf{u})$ moves with the sound-proof fluid
velocity $\mathbf{u}$. By Stokes' theorem, these sound-proof equations
generate circulation of $(\mathbf{u}+\mathbf{R})$ around
$\gamma_t(\mathbf{u})$ whenever the vectors $\nabla \theta$, $\nabla p'$ and $\nabla z$ are not collinear. Using advection of $\theta$
and the SP continuity equation, one finds conservation of
potential vorticity $q_{\rm SP}$ on fluid parcels, cf. equation
(\ref{pv-cons-EP}),
\begin{align} \label{pv-cons-SP}
\frac{\partial{q}_{\rm SP}}{\partial{t}}
+\mathbf{u}\cdot\nabla{q}_{\rm SP} &= 0\,,
\\  \nonumber
\hbox{where}\quad
{q}_{\rm SP} &= \frac{1}{D}
\nabla\theta\cdot{\rm curl}\,(\mathbf{u}+\mathbf{R})\,.
\end{align}
Consequently, the following domain integrated quantities are
conserved, for any smooth integrable function $\Phi$, cf. equation (\ref{EP-Casimirs}),
\begin{equation} \label{SP-Casimirs}
C_{\Phi} = \int 
D\,\Phi(\theta,q_{\rm SP})\,d^{\,3}x\,,
\quad\hbox{for all }\,\Phi\,.
\end{equation}
%

%%%%%%%%%%%%%%%%%%%%%%%%%%%%%%%%%%%%%%%%%%%%%%%%%%%%%%%%%%%%%%%%%%%%
\subsection{Energy conservation, Lie-Poisson Hamiltonian
formulation and nonlinear Lyapunov stability analysis} 

One uses the Legendre transform \eqref{EP-erg} of the Lagrangian (\ref{SP-lag}) to find the corresponding Hamiltonian,
\begin{align}\nonumber
H_{\SP} &= \int 
\bigg(\frac{1}{2D} |\mathbf{m}-D\mathbf{R}|^2
+ Dgz
+ C_v\pi_0(z)\,D\,\theta\bigg) \\
&  \qquad - \,
{p'}\!
\underbrace{
\Big(
\rho_0(z)\varTheta(\theta_0(z)) - D\,\varTheta(\theta)\Big)
}_{\hbox{SP constraint}}
d^{\,3}x
\,.\label{SP-LP-Ham}
\end{align}

The Lie-Poisson bracket (\ref{LPB}) now generates the SP motion 
equation \eqref{SP-mot-eqn1}, as well as  the auxillary equations \eqref{theta-advect}--\eqref{D-cont} and the SP constraint \eqref{SP-constraint2} from the Hamiltonian $H_{\SP}$ according to equation (\ref{mu-dot-syst}).
The conserved energy is obtained by evaluating the Hamiltonian $H_{\SP}$ on the SP constraint, as
\begin{equation} \label{sp-erg}
E_{\SP} = \int 
\bigg(\frac{1}{2} |\mathbf{u}|^2
+ gz
+ C_v\pi_0(z)\theta\bigg)
D \,d^{\,3}x
\,.
\end{equation}
This is the sum of the kinetic energy, gravitational potential energy and thermodynamic internal energy. That is, the SP constraint \eqref{SP-constraint2} does not interfere with conservation of the usual total energy.

\paragraph{Variational principle for SP equilibrium solutions.}
The equilibrium solutions of the SP equations occur at
critical points of the sum
$H_{\Phi}$, where
\begin{equation} \label{crt-pt}
H_{\Phi} = H_{\SP} + C_{\Phi}
\,,
\end{equation}
and $C_{\Phi}$ represents a family of Casimirs given by
\begin{equation} \label{casimir-def}
C_{\Phi} = \int 
D\,\Phi(\theta,q)\,d^{\,3}x\,,
\quad\forall\,\Phi\,.
\end{equation}
Here $\Phi$ is an arbitrary smooth function and $q$ is defined upon using the definition of the momentum density $\mathbf{m}$ in \eqref{m-def} as, 
\[
{q}\equiv\frac{1}{D}\nabla{\theta}\cdot{\rm curl}
\big(\mathbf{m}/D\big).
\]
Thus, the Casimir conservation laws for our Lie-Poisson Hamiltonian
formulation of the three-dimensional anelastic equations in the
Eulerian fluid description provide a constrained-energy variational
principle for the equilibrium solutions of the dynamical SP
equations and form a basis for determining their Lyapunov stability
conditions, as done for the Euler-Boussinesq equations in \cite{AHMR1986}. 
The AA and PIA equations comprise special cases of the SP equations when 
$\varTheta(\theta)=1-\alpha + \alpha \theta$. 
In turn, the Euler-Boussinesq equations form a special case of the AA and PIA equations in which the base state is constant. Consequently, the analyses of the nonlinear Lyapunov
stability conditions for the equilibrium solutions of the three-dimensional SP equations and their special cases the AA and PIA equations all follow a similar procedure to
that performed in \cite{AHMR1986}, which will produce a
similar result, modified according to the non-constant base state.

%%%%%%%%%%%%%%%%%%%%%%%%%%%%%%%%%%%%%%%%%%%%%%%%%%%%%%%%%%%%
\section{Sound-proof vertical slice models}\label{vertsliceAA+PIA-sec}

Slice models are used to describe the formation of fronts
in the atmosphere and ocean, modelling the situation in
which there is a strong horizontal temperature gradient
which maintains a vertical shear flow in the direction
tangential to the gradient, through geostrophic balance. On
the f -plane, this base flow can be modelled with a three-
dimensional flow with constant temperature gradient in the
y-direction, and velocity pointing in the x-direction with a
linear shear in the z-direction. This base flow is unstable
to y-independent perturbations to all three components of
velocity and temperature, rapidly leading to y-independent
front formation. The y-component of velocity is coupled
to the other variables through the constant y-gradient
of temperature. In the case of the incompressible Euler-
Boussinesq equations, solutions of these equations are also
solutions of the full three-dimensional equations. In other
cases, solutions of the three-dimensional equations are not
recovered but the slice models provide very useful tools for
analysing numerical methods for the atmosphere since they
can be compared with semigeostrophic solutions {}\cite {Cu2007}.

{{}}\cite {CoHo2012} considered variational formulations for vertical slice
models in which the prognostic variables are independent of $y$, with
the exception of potential temperature, which has a time-independent 
constant $y$-derivative, \emph{i.e.},
\begin{align}
\theta(x,y,z,t) = \theta_S(x,z,t) + (y-y_0){s},
\end{align}
where $s$ is a constant.  Here, we adopt the notation
$\mathbf{u}=(u,w)$, and treat $v$ separately, where $u$, $v$ and $w$
are the $x$- $y$- and $z$-components of velocity, respectively. We
also write $\nabla=(\partial_x,\partial_z)$. Consequently, the
$y$-independent mass density $D(x,z,t)$ satisfies
\begin{align}
\partial_tD + \nabla\cdot(\mathbf{u}D) = 0
\,
\label{D-eqn},
\end{align}
and three-dimensional scalar tracer equation \eqref{theta-advect}
becomes a dynamic equation for 
$\theta_S(x,z,t)$ which satisfies,
\begin{align}
\partial_t\theta_S + \mathbf{u}\cdot\nabla\theta_S + v{s} =0
\,.
\label{theta-eqn}
\end{align}

For a slice Lagrangian $l[\mathbf{u},v,D,\theta_S]$, \cite{CoHo2012}
showed that these conditions lead to the \emph{slice Euler-Poincar\'e
equations} (written here in curl form),
\begin{align}
\begin{split}
\pp{}{t}\frac{1}{D}\dede{\ell}{\mathbf{u}} 
&- \mathbf{u}\times\left(\curl\frac{1}{D}\dede{\ell}{\mathbf{u}}\right) 
 \\ 
& +\
\nabla\left(
\mathbf{u}\cdot\frac{1}{D}\dede{\ell}{\mathbf{u}}-\dede{\ell}{D}\right) 
\\
& +\
 \frac{1}{D}\dede{\ell}{v}\nabla v +
\frac{1}{D}\dede{\ell}{\theta_S}\nabla\theta_S=0, 
\end{split}
\label{EPSD-u} 
\\ 
\pp{}{t}\frac{1}{D}\dede{\ell}{v}
&+ \mathbf{u}\cdot\nabla\frac{1}{D}\dede{\ell}{v}
+\frac{1}{D}\dede{\ell}{\theta_S}{s}  = 0\,,
\label{EPSD-v}
\end{align}
where in the slice notation we write
$\MM{u}\times\curl\MM{v}=(-w(\partial_zv_1-\partial_xv_3),u(\partial_zv_1-\partial_xv_3))$. The system (\ref{EPSD-u}--\ref{EPSD-v}) is
completed by including the advection equations (\ref{D-eqn}) and
(\ref{theta-eqn}) for $D$ and $\theta_S$, respectively.

\paragraph{Kelvin-Noether theorem.}
\cite{CoHo2012} showed that the equations (\ref{EPSD-u}--\ref{EPSD-v})
can be combined to obtain
\begin{eqnarray}
\nonumber 
\pp{}{t}\frac{1}{D}\overline{\MM{m}}
 + \mathbf{u}\times \curl\left(\frac{1}{D}\overline{\MM{m}}\right)
+ \nabla \left(\MM{u}\cdot\overline{\MM{m}}-\dede{\ell}{D}\right)&=&0,
\label{EPSD-curl}
\end{eqnarray}
where 
\[
\overline{\MM{m}} = s\dede{l}{\MM{u}} - \dede{l}{v}\nabla\theta_S.
\]
This leads to the Kelvin-Noether conservation law for circulation
obeyed by slice models in the form
\begin{align}
\frac{d}{dt}\oint_{c(\mathbf{u})}
\frac{\overline{\MM{m}}}{D}
\cdot\dx
=
0
\,,
\label{EPslice-Kel}
\end{align}
where $c(\mathbf{u})$ is \emph{any} closed material loop that moves
with the slice fluid velocity $\mathbf{u}$ in the vertical slice plane
(\emph{i.e.}, not in the $y$-direction).
\paragraph{Potential vorticity}

Equation \eqref{EPslice-Kel} implies that potential vorticity $q$ is
conserved along flow lines of the fluid velocity $\mathbf{u}$,
\begin{align}
\partial_t q + \mathbf{u}\cdot\nabla q = 0
\,,
\label{EPSPslice-PVcons}
\end{align}
with potential vorticity
\begin{equation}
q = \nabla^\perp\overline{\MM{m}}
 = 
s\nabla^\perp\cdot\left(\frac{1}{D}\dede{\ell}{\mathbf{u}}\right)
-\nabla^\perp\dede{l}{v}\cdot\nabla\theta_S,
\,
\label{EPSPslice-PV}
\end{equation}
which is in fact the usual Ertel PV written in the slice variable notation.
\paragraph{Energy conservation}
As for the three-dimensional case, the absence of explicit time
dependence in the slice Lagrangian $\ell\,[\mathbf{u},v,D,\theta_S\,]$
gives the {\bfi conserved domain integrated energy}, via Noether's
theorem for time translation invariance. This energy is again easily
calculated using the {\bfi Legendre transform} to be
\begin{align*}
E\,[\mathbf{u},v,D,\theta_S\,] =& \int 
\Big(\mathbf{u}\cdot
\frac{\delta \ell}{\delta \mathbf{u}}
+v\dede{\ell}{v}
\Big) \diff{x}\diff{z}\ 
\\
&\qquad  - \ell\,[\mathbf{u},v,D,\theta_S\,]
\,.
\end{align*}
When the Legendre transform is completed to express
$E\,[\mathbf{u},v,D,\theta_S\,]$ as $H\,[\mathbf{m},n,D,\theta_S\,]$
with $\mathbf{m}\equiv\delta \ell/\delta \mathbf{u}$,
${n}\equiv\delta \ell/\delta v$,
$\delta{H}/\delta\mathbf{m}=\mathbf{u}$, and $\delta{H}/\delta n=v$,
the Euler--Poincar\'e system
(\ref{D-eqn}--\ref{EPSD-v}) may be
  expressed in Hamiltonian form as
\begin{equation}
\frac{\partial \mu}{\partial t}=\{\mu,H\}\,,
\quad \hbox{with}\quad
\mu\in[\mathbf{m},D,\theta_S\,]\,,
\end{equation}
and the Lie-Poisson bracket is given by
\begin{align} \nonumber
& \{F,H\}[\mathbf{m},D,\theta\,] =
\\ \nonumber
& \ - \, \int \MM{m}\cdot\left(\left(\dede{F}{\MM{m}}\cdot\nabla\right)
\dede{H}{\MM{m}} - \left(\dede{H}{\MM{m}}\cdot\nabla\right)\dede{F}{\MM{m}}
\right)\diff{x}\diff{z} \\
\nonumber & -
\int n\left(
\dede{F}{\MM{m}}\cdot\nabla\dede{H}{n}
-\dede{H}{\MM{m}}\cdot\nabla\dede{F}{n}
\right)\diff{x}\diff{z} \\
\nonumber & - 
\int \left( \dede{F}{\MM{m}}\cdot\nabla\theta_S + \dede{F}{n}s\right)
\dede{H}{\theta_S}\diff{x}\diff{z} \\
\nonumber & + 
\int \left( \dede{H}{\MM{m}}\cdot\nabla\theta_S + \dede{H}{n}s\right)
\dede{F}{\theta_S}\diff{x}\diff{z}.
\end{align}

%%%%%%%%%%%%%%%%%%%%%%%%%%%%%%%%%%%%%%%%%%%%%%%%%%%%%%%%%%%%
\subsection{Hamilton's principle for sound-proof slice models}
\label{SP-eqns-sec-slice}

In the present notation, the Lagrangian for the Sliced Sound-proof
Approximation (SSA) in Eulerian $(x,z)$ coordinates is,
%\begin{align}
%\begin{split}
%l_{\SAA}\big[\mathbf{u}, v,D,\theta_S\big] 
%&= 
%\int_{\Omega} \frac{D}{2}\left(|\mathbf{u}|^2 + v^2\right)
%+ fDv x
%- gDz
%\\
%&\hspace{1cm}- DC_v\pi_0(z)\theta_S 
% \\
%& \hspace{1cm} +p'\!\!
%\underbrace{\,
%\bigg(1-\frac{D}{\rho_0(z)}\bigg)
%}_{\hbox{AA constraint}}
%\diff{x}\diff{z}.
%\end{split}
%\label{SAA-Lag}
%\end{align}
%
%The corresponding Lagrangian for the Sliced Pseudo-Incompressible
%Approximation (SPIA) is
\begin{align}
l_{\SSA}\big[\mathbf{u}, v,D,\theta_S\big] 
&= 
\int_{\Omega} \frac{D}{2}\left(|\mathbf{u}|^2 + v^2\right)
+ fDv x
- gDz
\nonumber\\
&\hspace{1cm}- DC_v\pi_0(z)\theta_S 
\label{SSA-Lag}\\
& \hspace{-5mm} + p'\!
\underbrace{\,
\Big(
 \rho_0(z)\varTheta(\theta_S^0(z)) - D\,\varTheta(\theta_S)\Big)
}_{\hbox{SSA constraint}}\,\diff{x}\diff{z},
\nonumber
\end{align}
As before, the constraints are imposed by the
pressure as a Lagrange multiplier.
%

%except for one term. For the SAA Lagrangian,
%\begin{align}
%\begin{split}
%\frac1D\dede{l_{\SAA}}{\mathbf{u}}  &=  \mathbf{u}
%\,, \qquad \frac1D\dede{l_{\SAA}}{v}  =  v + fx
%\,, \\
%\dede{l_{\SAA}}{D} & =  \frac{1}{2}\left(|\mathbf{u}|^2 + v^2\right) + fvx - gz - C_v\pi_0(z)\theta_S
%- \frac{p'}{\rho_0(z)}
%\,, \\ &  \frac1D\dede{l_{\SAA}}{\theta_S}= -\, C_v\pi_0(z)
%\,,
%\end{split}
%\label{SAA-Lag-deriv}
%\end{align}
 
The variational derivatives of the SSA Lagrangian are,
\begin{align}
\begin{split}
 \frac1D\dede{l_{\SSA}}{\mathbf{u}}  &=  \mathbf{u}
\,, \qquad
 \frac1D\dede{l_{\SSA}}{v}  =  v + fx
\,, \\
\dede{l_{\SSA}}{D} & =  \frac{1}{2}\left(|\mathbf{u}|^2 + v^2\right) + fvx \\&\hspace{5mm}
- gz - C_v\pi_0(z)\theta_S 
- p'\varTheta(\theta_S)
\,, \\
 \frac1D\dede{l_{\SSA}}{\theta_S}
&= -\, C_v\pi_0(z) -p'\varTheta'(\theta_S)
\,.
\end{split}
\label{SPIA-Lag-deriv}
\end{align}
For the sliced sound-proof approximation SSA, the Euler-Poincar\'e approach
yields the following equations on the slice semidirect product with
advected density $D$ and scalar $\theta_S$:
\begin{align}
\begin{split}
&\pp{}{t}\mathbf{u} + \mathbf{u}\cdot\nabla \mathbf{u} 
 -fv\hat{\MM{x}} 
 \\
 &\quad
 + \left(g + C_v\dd{\pi_0}{z}\theta_S\right)\hat{\MM{z}} 
 + \varTheta(\theta_S)\nabla p'
=0, \\&
\pp{v}{t}
+ \mathbf{u}\cdot\nabla v + \hat{\MM{x}}\cdot\MM{u}f 
 \\ &
\quad 
-(C_v\pi_0(z)+p'\varTheta'(\theta_S))s 
 = 0\,, \\ &
\pp{D}{t} + \nabla\cdot(\MM{u}D)  =  0\,\ 
\\ &
\pp{\theta_S}{t} + \MM{u}\cdot\nabla\theta_S + vs 
 =  0,
\end{split}
\end{align}
For $\varTheta(\theta)=1-\alpha + \alpha \theta$, we have SPIA flows for $\alpha=1$ and SAA flows for $\alpha=0$. For $\alpha=0$ and $\rho_0(z)=const$, one recovers sliced  incompressible flows. Note that the solutions of the slice equations for SSA models are
not solutions of the full three-dimensional equations.

One obtains the
expected Kelvin circulation conservation law \eqref{EPslice-Kel},
written for the SSA models as
\[
\frac{d}{dt}\oint_{c(\mathbf{u})}\hspace{-2mm} 
\left(s\mathbf{u}  - (v+fx) \nabla \theta_S  \right)\cdot \dx
=
0
\,,
\]
which leads to potential vorticity conservation
\[
\partial_t q + \mathbf{u}\cdot\nabla q = 0
\]
with potential vorticity
\[
q = s\nabla^\perp\cdot\mathbf{u}
-\nabla^\perp(v+fx)\cdot\nabla\theta_S,
\]
This is the same circulation theorem as for the Euler-Boussinesq slice
model in \cite{CoHo2012}. 

The Euler-Poincar\'e slice equations are Hamiltonian, with conserved
energy
\begin{align*}
H = \int_{\Omega} \frac{1}{2D}&\left(|\mathbf{m}|^2 + (n-Dfx)^2\right)
\\&+
gDz + DC_v\pi_0(z)\theta_S \,d^{\,3}x
\,,
\end{align*}
and Casimir conservation laws,
 \begin{align}
 C_\Phi = \int D\Phi(q)\,\diff{V}
 \,,
\end{align}
for an arbitrary smooth function $\Phi$.
Furthermore, critical points of the sum 
\begin{align}
H_C = H + C_\Phi 
\end{align}
are equilibrium solutions. This is a general feature of Lie-Poisson
Hamiltonian theories, see, e.g., \cite{HoMaRaWe1985}, and in this case is the basis for studying Lyapunov stability for critical-point
equilibria of the slice models. However, this
feature will not be pursued further in the present paper.

%%%%%%%%%%%%%%%%%%%%%%%%%%%%%%%%%%%%%%%%%%%%%%%%%%%%%%%%%%%%%%
\section{Conclusions}\label{conc-sec}

In this paper we derived in the Euler-Poincar\'e variational framework a family of 3D soundproof models that contains the Lipps-Hemler anelastic approximation (AA) and the Durran pseudo-incompressible approximation (PIA) as special cases.  
In this family of 3D soundproof models, the pressure arises as a Lagrange multiplier enforcing a constraint on the density, just as in the standard incompressible case.
The models in the sound-proof family only differ in how the constraint is enforced. 
We have explained how these constraint forces relate to the standard forms of the equations. 
Having found Lagrangians that lead to these models
in Euler-Poincar\'e form, we immediately deduce that the models have a
conserved energy, inherit a Lie-Poisson structure, and have a
Kelvin-Noether circulation theorem and corresponding conserved
potential vorticity. This formulation opens the way for further
approximations in the Lagrangian, such as Lagrangian averaging or
balanced models. We also
briefly mentioned the interesting case of time-dependent
background profiles, which was discussed in \cite{Du2008}.
These models are useful for driving local area models using
profiles from global model output, or to model seasonal
variations or regional climate change. In this case, potential
vorticity will be conserved but we expect changes in
domain-integrated energy in general, since there is now an
explicit time-dependence in the Lagrangian. It is interesting
to consider the case where the time-dependence in the
background flow is slow, since in that case we expect the
energy to be almost invariant; we will study this problem in
future work.

In the second half of the paper, we adapted this
derivation to obtain vertical slice soundproof models. These vertical
slice models are used to model front formation, and allow for full
three dimensional velocity vectors, but with all fields being
$y$-dependent except for potential temperature which is assumed to
have a constant gradient in the $y$-direction. Unlike the
incompressible Boussinesq model, our slice models do not recover
solutions of the full 3D equations since they have
linearised the equation of state under the assumption of small
$\theta$ variation in the $y$-direction. However, we believe that they
can play a role in the validation of numerical schemes and closures as
part of the programme of asymptotic limit analysis described in
\cite{Cu2007}, in which numerical solutions are compared with
numerical solutions of the limiting semi-geostrophic model, which are
most easily obtained for models in slice configuration using a
geometric algorithm. Initial work on this programme for Boussinesq
models is shown in \cite{ViCoCu2013}. A requirement for such tests is
that the model being tested has a conserved energy and potential
vorticity. Both conservation laws are provided by these models, which can be obtained
from minor adaptations of existing soundproof codes.

%%%%%%%%%%%%%%%%%%%%%%%%%%%%%%%%%%%%%%%%%%%%%%%%%%%%%%%%%%%%%%%%%%%%
\subsection*{Acknowledgements}
We thank P. Smolarkiewicz for stimulating conversations about this work and for providing helpful comments on a much earlier draft. 
We also thank D. Durran for discussions about sound-proof models. 
The authors are also
grateful to M. J. P. Cullen and A. Visram for very useful and
interesting discussions about slice models. The work by CJC was
partially supported by the Natural Environment Research Council Next
Generation Weather and Climate programme. The work by DDH was
partially supported by Advanced Grant 267382 from the European
Research Council.

\bibliography{Soundproof}

\begin{thebibliography}{HMRW85}

\bibitem[AHMR86]{AHMR1986}
H.D.I. Abarbanel, D.~D. Holm, J.~E. Marsden, and T.~Ratiu.
\newblock Nonlinear stability analysis of stratified ideal fluid equilibria.
\newblock {\em Phil Trans. Roy. Soc.}, 318:349--409, 1986.

\bibitem[Ban95]{Ba1995}
P.~R. Bannon.
\newblock Hydrostatic adjustment: {L}amb's problem.
\newblock {\em J. Atmos. Sci.}, 52:1743--1752, 1995.

\bibitem[Ban96]{Ba1996}
P.~R. Bannon.
\newblock On the anelastic approximation for a compressible atmosphere.
\newblock {\em J. Atmos. Sci.}, 53:3618--3628, 1996.

\bibitem[Ber95]{Be1995}
P.~Bernardet.
\newblock The pressure term in the anelastic model: a symmetric elliptic solver
  for an {Arakawa C} grid in generalized coordinates.
\newblock {\em Monthly Weather Rev.}, 123:2474--2490, 1995.

\bibitem[CH12]{CoHo2012b}
C.J. Cotter and D.D. Holm.
\newblock On {Noether's Theorem for the Euler--Poincar{\'e}} equation on the
  diffeomorphism group with advected quantities.
\newblock {\em Foundations of Computational Mathematics}, pages 1--21, 2012.

\bibitem[CH13]{CoHo2012}
C.~J. Cotter and D.~D. Holm.
\newblock A variational formulation of vertical slice models.
\newblock To appear in {\it Proc Roy Soc A}, preprint available at
  http://arxiv.org/abs/1211.2067, 2013.

\bibitem[Cul07]{Cu2007}
M.J.P. Cullen.
\newblock Modelling atmospheric flows.
\newblock {\em Acta Numerica}, 2007.

\bibitem[DF69]{DF1969}
D.~R. Dutton and G.~H. Fichtl.
\newblock Approximate equations of motion for gases and liquids.
\newblock {\em J. Atmos. Sci.}, 1969.

\bibitem[Dur89]{Du1989}
D.~R Durran.
\newblock Improving the anelastic approximation.
\newblock {\em J. Atmos. Sci.}, 46:1453--1461, 1989.

\bibitem[Dur08]{Du2008}
D.~R. Durran.
\newblock A physically motivated approach for filtering acoustic waves from the
  equations governing compressible stratified flow.
\newblock {\em J. Fluid Mech.}, 601:365--379, 2008.

\bibitem[HMR98]{HMR1998a}
D.~D. Holm, J.~E. Marsden, and T.~Ratiu.
\newblock The {Euler-Poincar\'e equations} and semidirect products with
  applications to continuum theories.
\newblock {\em Adv. in Math.}, 137, 1998.

\bibitem[HMR02]{HMR1998b}
D.~D. Holm, J.~E. Marsden, and T.~S. Ratiu.
\newblock The {Euler--Poincar\'{e} equations} in geophysical fluid dynamics.
\newblock In {\em Proceedings of the Isaac Newton Institute Programme on the
  Mathematics of Atmospheric and Ocean Dynamics}. Cambridge University Press,
  2002.

\bibitem[HMRW85]{HoMaRaWe1985}
D.~D. Holm, J.~E. Marsden, T.~S. Ratiu, and A~Weinstein.
\newblock Nonlinear stability of fluid and plasma equilibria.
\newblock {\em Physics Reports}, 123:1--116, 1985.

\bibitem[Kle09]{Kl2009}
R.~Klein.
\newblock Asymptotics, structure, and integration of soundproof atmospheric
  flow equations.
\newblock {\em Theor. Comput. Fluid Dyn.}, 23:161--195, 2009.

\bibitem[LH82]{LH1982}
F.~B. Lipps and R.~S. Hemler.
\newblock A scale analysis of deep moist convection and some related numerical
  calculations.
\newblock {\em J. Atmos. Sci.}, 39:2192--2210, 1982.

\bibitem[MR95]{MR1994}
J.~E. Marsden and T.~S. Ratiu.
\newblock {\em Introduction to Mechanics and Symmetry}, volume~17 of {\em Texts
  in Applied Mathematics}.
\newblock Springer-Verlag, 1995.

\bibitem[OP62]{OgPh1962}
Y.~Ogura and N.~A. Phillips.
\newblock Scale analysis for deep and shallow convection in the atmosphere.
\newblock {\em J. Atmos. Sci.}, 19:173--179, 1962.

\bibitem[SS92]{SS1992}
J.~F. Scinocca and T.~G. Shepherd.
\newblock Nonlinear wave-activity conservation laws and {Hamiltonian} structure
  for the two-dimensional anelastic equations.
\newblock {\em J. Atmos. Sci.}, 49:5--27, 1992.

\bibitem[Thu08]{Th2008}
J.~Thuburn.
\newblock Some conservation issues for the dynamical cores of {NWP} and climate
  models.
\newblock {\em Journal of Computational Physics}, 227(7):3715--3730, 2008.

\bibitem[VCC13]{ViCoCu2013}
A.R. Visram, C.J. Cotter, and M.J.P. Cullen.
\newblock A framework for evaluating model error using asymptotic convergence
  in the {Eady} model.
\newblock Submitted, 2013.

\end{thebibliography}

\end{document}